\documentclass[12pt]{article}
\usepackage{alltt,graphicx,amssymb}
\usepackage{amsmath,mathrsfs,color}
\usepackage{caption,subcaption}
\usepackage[latin1]{inputenc}
\usepackage{authblk}
\usepackage[left=3cm]{geometry}
\newcommand{\ba}{\begin{eqnarray}}
\newcommand{\ea}{\end{eqnarray}}
\renewcommand{\rho}{\varrho}
\newcommand{\qo}{q_{0}}
\newcommand{\qb}{\bar{q}}
\newcommand{\tho}{\theta_0}

\def\pt{{\partial}}
\def\ev{{\bf e}}
\def\s0{{s_0}}

\def\nv{{\bf n}}
\def\mv{{\bf m}}

\def\tv{{\bf t}}
\def\fv{{\bf f}}

\def\tv{{\bf t}}

\def\F{{\rm F}}
\def\E{{\rm E}}

\def\Fv{{\bf F}}

\def\Mv{{\bf M}}

\def\Tv{{\bf T}}

\def\bs{\bar{s}}
\def\bt{\bar{\theta}}

\def\t0{{\theta_0}}

\def\de{\delta}

\def\eps{\epsilon}
\def\d{{\rm d}}

\newcommand{\vc}{v_0^{_{(1)}}}
\newcommand{\va}{v_0^{_{(\infty)}}}

\newcommand{\bGa}{ \mbox{ \hspace{-1mm}\boldmath $\Gamma$ \hspace{-1mm}}}

\newcommand{\bphi}{ \mbox{ \hspace{-1mm}\boldmath $\phi$ \hspace{-1mm}}}
\newcommand{\bpsi}{ \mbox{ \hspace{-1mm}\boldmath $\psi$ \hspace{-1mm}}}

\title{\textbf{Growth-induced blisters in a circular tube}}

\author[1]{R. De Pascalis}
\author[2]{G. Napoli%
\thanks{\texttt{gaetano.napoli@unisalento.it}}}
\author[3]{S. S. Turzi}

\affil[1]{\small\!School of Mathematics, University of Manchester, Manchester M13 9PL, UK.}
\affil[2]{Dipartimento di Ingegneria dell'Innovazione, Universit\`a del Salento. Via per Monteroni Edificio ``Corpo O'', 73100, Lecce, Italy.}
\affil[3]{Dipartimento di Matematica, Politecnico di Milano, Piazza Leonardo da Vinci 32, 20132 Milano, Italy.}

\date{27/11/2013}

\begin{document}
\thispagestyle{plain}

\maketitle

\begin{abstract}
The growth of an elastic film adhered to a confining substrate might lead to the formation of delimitation blisters. Many results have been derived when the substrate is flat. The equilibrium shapes, beyond small deformations, are determined by the interplay between the sheet elastic energy and the adhesive potential due to capillarity. Here, we study a non-trivial generalization to this problem and consider the adhesion of a growing elastic loop to a confining \emph{circular} substrate. The fundamental equations, i.e., the Euler Elastica equation, the boundary conditions and the transversality condition, are derived from a variational procedure. In contrast to the planar case, the curvature of the delimiting wall appears in the transversality condition, thus acting as a further source of adhesion. We provide the analytic solution to the problem under study in terms of elliptic integrals and perform the numerical and the asymptotic analysis of the characteristic lengths of the blister. Finally, and 
in contrast to previous studies, we also discuss the mechanics and the internal stresses in the case of vanishing adhesion. Specifically, we give a theoretical explanation to the observed divergence of the mean pressure exerted by the strip on the container in the limit of small excess-length.
\end{abstract}

% \begin{keyword}
% Constrained Elastica, Adhesion, Delamination, Growth
% \end{keyword}

\section{Introduction}
The classical theory of bending, due to Bernoulli and Euler more than four centuries ago, is still considered a key simplified model for understanding the mechanics of many {\it hard} and {\it soft systems}. Within this theory, the mechanical properties and the shape of rods and sheets can be determined by solving an ordinary differential equation: the fundamental equation of Euler's {\it Elastica}. Several problems may be tackled by this method, albeit with some variations, as for instance the occurrence of delamination blisters \cite{Williams:1997, Wagner:2013}, the adhesion of lipid tubules \cite{Rosso:1998}, growth mechanisms in climbing plants \cite{Goriely:2006}, the mechanics of the insertion of a guidewire into the artery of a patient \cite{Chen:2007}, the equilibria of the uplifted heavy elastic strip \cite{Domokos:2003} and the pattern formation of flexible structures induced by tight packing \cite{Boue:2006}.

In this paper, we  analyze the growth of a closed planar Euler-Bernoulli strip confined by a rigid \emph{circular} domain. We employ a very simple  growth mechanism and posit that the total length of the strip may be changed arbitrarily by a suitable external action. Thus, mathematically, we consider the total length of the strip as an adjustable parameter, whose governing equation has no need to be specified. Furthermore, we assume the strip to be inextensible and always at equilibrium. A simple rudimentary experimental setup can help to describe the physical phenomenon we wish to analyze. Let us imagine a flexible cylinder made out of a piece of paper, simply by gluing together the edges of a rectangular sheet.  Next, we insert this flexible cylinder into a rigid circular tube of smaller radius. The shape of the confined sheet, unavoidably, exhibits {\it blisters}, {\it i.e.}, regions of the sheet which are not in contact with the substrate but form inward protuberances. Even in the presence of an ideal 
frictionless substrate, part of the strip adheres to the confining wall as the circular geometry acts as an adhesion mechanism. 

An increase of adhesion may be further promoted by capillarity. Generally, adhesion by capillarity may occur in an elastic structure when its restoring ability is unable  fully to   overcome the interfacial attraction induced by liquid surface tensions. Various capillary adhesion phenomena can be observed at small scale in both natural phenomena  and industrial processes. For a more exhaustive overview of these topics, we refer the reader to the recent review article \cite{Liu:2012} and references therein.

The problem of a growing Elastica confined by a frictionless rigid circumference has been studied numerically in \cite{Boue:2006} in order to explain the packing of a flexible sheet inside a cylindrical cavity. This analysis has been subsequently extended to the growth of an Helfrich's membrane confined within a spherical domain \cite{Kahraman:2012}. In addition to the elastic problem, both these papers consider the complicated conditions arising from the contact with the container and from the self-contact. Other studies exploit the theory of elliptic integrals to characterize the adhesion of lipid tubules on curved substrates \cite{Rosso:1998}, the stability of clamped  elastic arches \cite{Patricio:1998} or the Euler buckling of constrained strait beams \cite{Domokos:1997}. More recently \cite{Wagner:2013}, the Elastica theory has been used to study the deformation of thin elastic sheets that adhere  to a stiff flat substrate by means of a surface potential. The authors use a 
combination of numerical and asymptotic techniques to predict the equilibrium shapes of this {\it sticky Elastica}. 

Our paper generalizes the theoretical results relative to the formation of delamination blisters on a flat surface, as reported by Wagner and Vella \cite{Wagner:2013}, to the case of a circular substrate. In \S 2, we obtain the equilibrium equations as extremal points of the energy functional subject to suitable constraints. The variational procedure is here slightly complicated by the fact that the end-points of the energy functional are not fixed, but are part of the unknowns. This entails that, besides the equilibrium equation and its boundary conditions, a further boundary condition is needed to determine the location of the detachment points. In contrast to the case of a flat substrate, the morphology of the strip is affected not only by the {\it elastocapillarity length}, but also by the container radius. In \S 3, the symmetrical equilibrium configurations of the strip are investigated using both an integral formulation and an asymptotic approach. Relevant measurable quantities (the blister height, the 
length of the adherent part and the internal forces) are provided as functions of the total length and the adherence strength.  Furthermore, we provide an asymptotic expansion of the solution in terms of a dimensionless parameter measuring the excess length of the beam with respect to the container length. Finally, in \S 4, we analyze the case of adherence due to the sole curvature when the delimiting wall is a frictionless unilateral contact. In this case, the tensional state of the entire strip and the forces exerted on the wall can be easily computed. As we shall see, these forces show a singular behaviour and thus lead to non-trivial conclusions.
\section{Variational problem}
In this section we derive the equilibrium equation and the boundary conditions that have to be fulfilled by the free part of the Euler-Bernoulli beam. Let us describe the geometry of the beam with a planar curve $\gamma$.  In the plane of the curve, we introduce a Cartesian frame of reference $(O;\ev_x,\ev_y)$, where $O$ is the origin and $\ev_x$, $\ev_y$ are the unit vectors along, respectively, the $x$ and the $y$ axes. Let $s$ be the arc-length along the curve and $\theta(s)$ the inflection angle. More precisely, $\theta(s)$ measures the anti-clockwise angle between $\ev_x$ and the tangent to the curve $\tv(s)$. Therefore, the Frenet curvature is $\theta_s(s)$, where the subscript denotes differentiation with respect to its argument. Each point $p$ on $\gamma$ can be parametrized by the Cartesian coordinates $x(s)$ and $y(s)$, so that its position vector is $(p - O) = x(s) \ev_x + y(s) \ev_y$. On the other hand, since $\tv = \d (p-O)/\d s$, it follows that
\begin{equation}
x_s = \cos \theta, \qquad y_s = \sin \theta. 
\label{para}
\end{equation}
Furthermore, we posit the following classical bending energy
\begin{equation}
W_b[\theta_s] =\frac{\kappa}{2}\int_{-\frac{\ell}{2}}^{\frac{\ell}{2}} (\theta_s - c_0)^2 \d s,
\end{equation}
where  the constant $c_0$ accounts for a possible spontaneous curvature of the beam, $\kappa$ is the bending rigidity and $\ell$ is the total length.
\begin{figure}[ht]
\centerline{\includegraphics[width=0.6\textwidth]{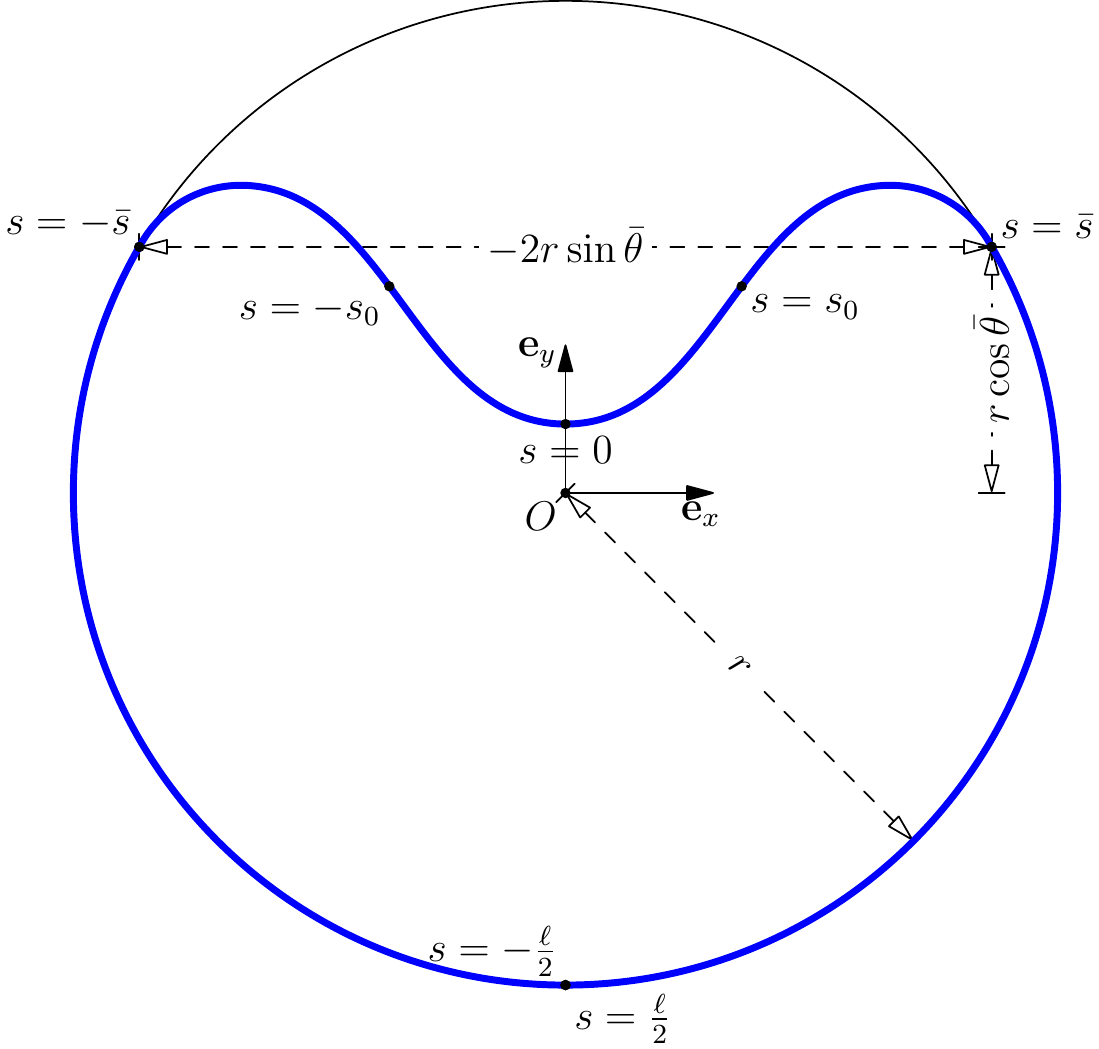}}
\caption{\small Schematic representation of the elastic strip, confined by a cylindrical wall of radius $r$. We assume that its shape has a mirror symmetry with respect to the $y$-axis, allowing the study only for the branch $s\ge0$. The free part of the curve is parametrized by values of the arc length in the range $s \in [0, \bs)$, where $\bs$ is the detachment point. At $s=s_0$ the curvature vanishes. The adopted conventions imply that the curvature is positive in $s\in[0,s_0)$ and negative in $s \in (s_0, \ell/2]$. In particular, $\theta_s \equiv -1/r$ throughout the adherent part.}
\label{fig1}
\end{figure}
It is worth noticing that the parametrization of $\gamma$ by means of the inflection angle $\theta$ automatically ensures the arc-length preservation, thus no Lagrange multiplier associated to the inextensibility constraint is needed. 
 
With reference to the schematic representation in Figure \ref{fig1}, we assume that the beam forms a unique blister. Thus, the beam at equilibrium consists of two parts: a free --non-adherent-- curve, described by $s \in (-\bs, \bs)$, and an adherent one, with $s$ in the range $s \in [-\ell/2,-\bs] \cup [\bs, \ell/2]$.  Hereinafter, we assume $\theta(s)$ odd, so that the problem can be studied in the interval $[0, \ell/2]$ only. 

In our simplified treatment there are only two detachment points, which correspond to the arc-lengths $-\bs$ and $\bs$, respectively and, of course, are constrained to lie along a circumference of radius $r$. A glance at Figure \ref{fig1} shows that their distance is $-2 r \sin \bt$, where $\bt$ is the value of the inflection angle at $\bs$ (note that $\bt \le 0$, by our conventions). Elementary geometric arguments yield also the following identity 
\begin{equation}
\bt := \theta(\bs) =  -\frac{\bs}{r}  + \frac{\ell - 2 \pi r}{2r}.
\label{sb:tb}
\end{equation}
On the other hand, the distance between the detachment points can be obtained by integrating \eqref{para}$_1$ in $[-\bs,\bs]$. As a consequence, the equilibrium solution should obey the global constraint
\begin{equation}
\int_{0}^{\bs} \cos \theta \d s = - r \sin\bt. 
\label{vinco}
\end{equation}
Thus, in the free region, the effective potential to minimize becomes
\begin{equation}
W_f[\theta, \theta_s;\bs] =\int_{0}^{\bs} \kappa(\theta_s - c_0)^2 \d s - 2T_x \left(r \sin\bt + \int_{0}^{\bs} \cos \theta \,\d s \right) ,
\end{equation}
where $T_x$ is a Lagrange multiplier and $\bs$ is to be determined in the minimization process. 
 
In the adherent region, the beam is in contact with the circular container and the bending energy is constant since $\theta_s = -1/r$. However, in order to account for elasto-capillarity effects, we further consider an adhesive potential describing the strip-substrate adhesion. We assume this in its simplest form taking it proportional to the length of the sticking region through a positive constant $w$, which is called the {\it adherence strength}. The energy associated to the adherent part is therefore
\begin{equation}
W_a[\bs] = \int_{\bs}^{\frac{\ell}{2}} \kappa\left(\frac{1}{r} + c_0\right)^2 \d s -2 \int_{\bs} ^{\frac{\ell}{2}} w \,\d s.
\end{equation}
Since the shape of the adherent part is fixed, this energy is a function of $\bs$ only.
\subsection{Euler-Lagrange equations and boundary conditions}
The equilibrium configurations are stationary points of the total free energy $W = W_f + W_a$. By adopting the notation as in \cite{Fomin:1963}, we consider two neighboring curves $\theta(s)$ and $\theta_h(s)$ such that
\begin{equation}
\theta_h(s) = \theta(s) +  h(s).
\end{equation}
The variational procedure must explicitly include the fact that the end points $s=0$ and $s= \ell/2$ are fixed, while the detachment point $s=\bs$ is not. Consequently, standard arguments \cite{Fomin:1963} show that the possible variations have to satisfy the following equations at the end-points and to first order
\begin{subequations}
\begin{equation}
\theta_h(0) = \theta(0), \qquad \theta_h(\ell/2) = \theta(\ell/2) , 
\end{equation}
\begin{equation}
 h(\bs) = \de \bt - \theta_s(\bs) \de \bs ,
\end{equation}
\end{subequations}
where $\de \bt := \theta_h(\bs+\de \bs)-\theta(\bs)$. Thus, by setting
\begin{equation}
g_f(\theta_s) =  \kappa(\theta_s - c_0)^2, \quad   g_c(\theta) = - 2T_x \cos \theta, \quad g_a = \kappa\left(\frac{1}{r} + c_0\right)^2  - 2 w ,
\label{gg}
\end{equation}
the first variation of $W$ is
\begin{eqnarray}
\de W = \int_0 ^{\bs}\left[\frac{\pt g_c}{\pt \theta} - \frac{\d }{\d s}\frac{\pt g_f}{\pt \theta_s} \right] h(s) \d s +
\left. \left(\frac{\pt g_f}{\pt \theta_s} - 2 T_x r \cos \theta \right) \right|_{s=\bs} \hspace{-2mm} \de \bt \nonumber \\
+ \left.\left(g_f  - \frac{\pt g_f}{\pt \theta_s} \theta_s  + g_c - g_a\right)\right|_{s=\bs}  \hspace{-2mm} \de \bs. 
\label{var1}
\end{eqnarray}
Since $\bs$ lies on a circumference of radius $r$ it follows that the variations $\de \bt$ and $\de \bs$ are not independent:
\begin{equation}
\de \bt = -\frac{\de \bs}{r} \, .
\label{vart}
\end{equation}
The substitution of equation \eqref{vart} into  \eqref{var1}, after some manipulations, yields
\begin{equation}
\de W = \int_0 ^{\bs}\left[\frac{\pt g_c}{\pt \theta} - \frac{\d }{\d s}\frac{\pt g_f}{\pt \theta_s} \right] h(s) \d s + \left[-\left(\frac{1}{r} +   \theta_s\right)\frac{\pt g_f}{\pt \theta_s} +   g_f   - g_a\right]_{s=\bs} \hspace{-2mm}\de \bs \, .
\label{var2}
\end{equation}
The equilibrium condition, $\de W = 0$, for any arbitrary choice of $h(s)$ and $\de \bs$, leads to the requirement that each term enclosed in square  brackets in \eqref{var2} must vanish. Therefore, once the explicit expressions of $g_a$, $g_c$ and $g_f$ as given in \eqref{gg} are taken into account, the following Euler-Lagrange equation is derived
\begin{equation}
\kappa \theta_{ss} - T_x \sin \theta = 0, \qquad s \in (0,\bs),
\label{pendolo}
\end{equation}
with the boundary conditions 
\begin{equation}
\theta(0) =0, \qquad \theta(\bs) = \bt. 
\label{compa}
\end{equation}
As expected, the angle $\theta(s)$ has to satisfy the non-linear pendulum equation. Nevertheless, contrary to the classic pendulum dynamics, the equation \eqref{pendolo} has to be solved with boundary conditions rather than initial conditions. Moreover, both the Lagrangian multiplier $-T_x$ (which plays the role of gravity in the pendulum analogy) and the boundary point $\bs$ are unknowns.  However, the vanishing of the coefficient of $\de \bs$ in \eqref{var2} gives a further condition at the detachment point: 
\begin{equation}
\kappa \left(\bt_s + \frac{1}{r}\right)^2 - 2 w = 0, 
\label{aderenza}
\end{equation}
which we refer to as the \emph{transversality condition}. This equation, along with Eqs.\eqref{vinco}, \eqref{pendolo} and \eqref{compa}, allows us to solve the problem and thus determine the unknowns $T_x$ and  $\bs$.

The condition \eqref{aderenza} is a special case of a more general \emph{adhesive condition} obtained in \cite{Rosso:1998}. It reflects the fact that there are two different sources of adhesion: (i) the adherence by curvature, which is proportional to the bending stiffness and is a decreasing function of the radius $r$; and (ii) the adhesive potential whose strength is provided by $w$. In the limit case where $w=0$, equation \eqref{aderenza} guarantees the continuity of the curvature $\theta_s$ at $s=\bs$. On the other hand, whenever the substrate is flat ($r \rightarrow \infty$), we correctly recover the adherence condition used in \cite{Wagner:2013}. 

By defining the {\it elasto-capillarity} as $\ell_{ec} = \sqrt{\kappa/w}$, \eqref{aderenza} reduces to
\begin{equation}
\bt_s = -\frac{1}{r} - \frac{\sqrt 2}{ \ell_{ec}}
\label{ad1}
\end{equation}
where the minus sign in front of $\sqrt 2/\ell_{ec}$ is due to the fact that the curvature radius at the detachment point cannot exceed that of the delimiting wall.
 
Finally, we remark that the spontaneous curvature $c_0$ plays no role in the equilibrium equations. Indeed, the energetic terms involving $c_0$ are null Lagrangians and, hence, they could possibly affect only the boundary conditions. However, since $\gamma$ is a closed curve, $c_0$  cannot have any effect on the equilibrium shape. %
\section{Equilibrium shapes}
We now examine a special class of equilibrium solutions, schematically shown in Figure \ref{fig1}. The expected equilibrium solution $\theta(s)$ is an increasing function for $s \in (0,\s0)$, while it decreases for $s \in (\s0, \bs)$. Let $\theta_0 = \theta(\s0) \in [0,\pi]$ be the maximum value of $\theta(s)$ in $(0,\bs)$. Standard arguments in the calculus of variations show that a first integral of \eqref{pendolo} is
\begin{equation}
\frac{1}{2}\theta^2_s = \tau(\cos \theta - \cos \theta_0),
\label{primo}
\end{equation}
where we have set $\tau = -T_x/\kappa$. To simplify the notation, we introduce 
\begin{equation}
\eta := \frac{1}{r} + \frac{\sqrt{2}}{ \ell_{ec}},
\end{equation}
and rewrite the transversality condition \eqref{ad1} as $\theta_s(\bs) = -\eta$. Therefore, Eq.\eqref{primo} yields 
\begin{equation}
\frac{1}{\tau} =   \frac{2}{\eta^2} (\cos \bt- \cos \t0 ),
\label{eq:tau}
\end{equation}
with $|\bt| \neq \t0$. By replacing \eqref{eq:tau} into \eqref{primo}, we finally deduce that
\begin{equation}
\theta_s =  \pm \eta \sqrt{\frac{\cos \theta - \cos \theta_0}{\cos \bt - \cos \theta_0}},
\label{tetas}
\end{equation}
where the sign $+$ (respectively, $-$) is to be used in the interval $s\in(0,\s0)$ (respectively, $s\in(\s0,\bs)$). By symmetry $\theta(0)=0$ and Eq.\eqref{tetas} evaluated at $s=0$ shows that $\cos\bt -\cos\t0 >0$. This gives a restriction on the possible values of $\bt$: $|\bt|<\t0$. Furthermore, \eqref{tetas} is an ordinary differential equation which can be solved by separation of variables in $(0,\bs)$. To this end, we change the variable of integration from $s$ to $\theta$
\begin{equation}
\d s = \pm \frac{1}{\eta} \sqrt{\frac{\cos \bt - \cos \theta_0} {\cos \theta - \cos \theta_0}} \, \d \theta
\end{equation}
and divide the integral into the two sub-regions where the function $\theta(s)$ is monotonic
\begin{equation}
\int_0^\t0 \frac{\d \theta}{\sqrt{\cos \theta - \cos \theta_0}}  -  \int_\t0^{\bt} \frac{\d \theta}{\sqrt{\cos \theta - \cos \theta_0}} =  \frac{\eta \bs}{ \sqrt{\cos \bt - \cos \theta_0}},
\label{sepa}
\end{equation}
where, on the right hand side, the boundary conditions \eqref{compa} have been used. Finally, equation \eqref{sepa} can be recast in the following form
\begin{equation}
4\F(q_0)-2\F(\bar q)  ={\eta \bs} {\sqrt \frac{1-\cos\theta_0}{\cos \bt - \cos \theta_0}},
\label{prima}
\end{equation}
where $\F$ denotes the incomplete elliptic integral of first kind \cite{Abramowitz:1970} and, for ease of notation, we set
\begin{equation}
q_0 := \left\{\frac{\t0}{2}, \csc^2 \frac{\t0}{2} \right\}, \qquad \bar q := \left\{\frac{\bt}{2}, \csc^2 \frac{\t0}{2} \right\}.
\end{equation}
Similarly, we reduce equation \eqref{vinco} as follows
\begin{equation}
\int_{0} ^{\t0} \frac{\cos \theta}{\sqrt{\cos \theta - \cos \theta_0}} \d \theta - \int_{\t0} ^{\bt} \frac{\cos \theta}{\sqrt{\cos \theta - \cos \theta_0}} \d \theta   = - \frac{ \eta r \sin \bar \theta}{ \sqrt{\cos \bt - \cos \theta_0}},
\label{pp}
\end{equation}
and rewrite the left hand side of \eqref{pp} in terms of elliptic integrals. With the aid of \eqref{prima}, we finally obtain
\begin{equation}
{2 \left(1-\cos\theta_0\right) \left[2   \E (q_0)-  \E (\bar q)\right]} = - \eta(\bs \cos \t0 + r\sin \bar \theta)\sqrt\frac{{1-\cos\theta_0}}{\cos \bt - \cos \theta_0},
\label{seconda}
\end{equation}
where $\E$ represents the incomplete elliptic integral of second kind \cite{Abramowitz:1970}.

By using equation \eqref{sb:tb},  we can eliminate $\bs$ in equations  \eqref{prima} and \eqref{seconda}  in favor of $\bt$. Thus, the solutions of the nonlinear transcendental equations \eqref{prima} and \eqref{seconda} (whenever exist) give the values of $\bt$ and $\t0$ as functions of the length $\ell$, the elasto-capillarity length $\ell_{ec}$ and the radius $r$. Hence, the solution is completely determined.

\begin{figure}[ht]
\centerline{\includegraphics[width=0.5\textwidth]{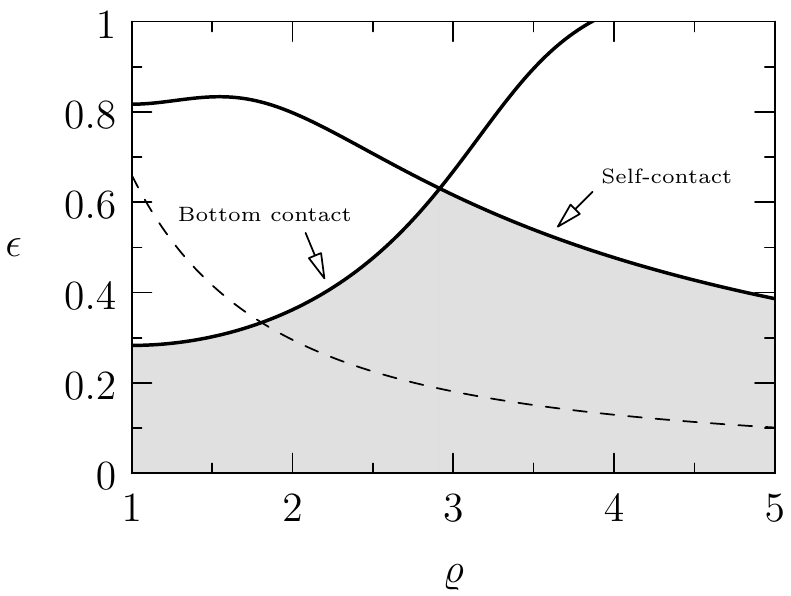}}
\caption{\small The shaded area represents the region of the parameters $\eps$ and $\varrho$ where the solution is valid. This region is delimited by two solid curves. The first (black line) identifies the configuration at which the blister vertex ($s=0$) is in contact with the diametrically opposite point $s=\ell/2$. We say that the elastic strip ``touches the container at the bottom''. For values of $(\varrho,\eps)$ along the second curve (blue line), the elastic strip is in contact with itself in an intermediate point (self-contact). Below the dashed line, $\ell^{(adh)}$ is a decreasing function of $\eps$, while above it is an increasing function of $\eps$. }
\label{fig:limits}
\end{figure}

It is now of special interest to study the expression of the length of the adherent portion of the strip, defined as $\ell^{(adh)} := \ell - 2 \bs$, and that of the blister height, defined as $\delta:=r-y(0)$. These are, in fact, quantities easily accessible experimentally. The former is simply given by
\begin{equation}
\ell^{(adh)}  = 2r(\pi + \bt) \, . 
\label{eq:laderenza}
\end{equation}
In order to derive $\delta$ as a function of $\bt$ and $\tho$, we note that $\delta = r-(y(0)-y(\bs))-y(\bs)$, so that we can write
\begin{equation}
\begin{split}
\delta & = r(1-\cos\bt) + \int_{0}^{\bs}\sin\theta(s) \d s 
= r(1-\cos\bt) -\frac{1}{\tau} \int_{0}^{\bs}\theta_{ss} \d s \\
& = r(1-\cos\bt) -\frac{1}{\tau} \left[\theta_s(\bs) - \theta_s(0) \right] \, . \nonumber
\end{split}
\end{equation}
We then use Eqs.\eqref{eq:tau} and \eqref{tetas} to simplify further and obtain
\begin{equation}
\delta = r(1-\cos\bt) + \frac{2}{\eta}\sqrt{\cos\bt-\cos\tho}\Big(\sqrt{\cos\bt-\cos\tho}+\sqrt{1-\cos\tho} \Big) \, .
\label{eq:blister_height}
\end{equation}
The solution in terms of elliptic integrals is relatively simple to implement computationally. However, the type of solution we seek remains valid as long as there is no self-intersection and the strip does not touch the lower part of the circular container. For later convenience, it is apposite to introduce the following adimensional quantities
\begin{equation}
\eps = \frac{\ell - 2\pi r}{2\pi r} \, , \qquad \varrho=\eta r = 1 + \frac{\sqrt{2}\,r}{\ell_{ec}}\, .
\label{eq:epsilon_rho}
\end{equation}
The former measures the excess length with respect to the confining circumference, while the latter determines the relative importance of the adhesion induced by curvature with respect to the adhesion by elasto-capillarity.

\begin{figure}[ht]
\centering
\begin{subfigure}{0.35\textwidth}
\includegraphics[width=\textwidth]{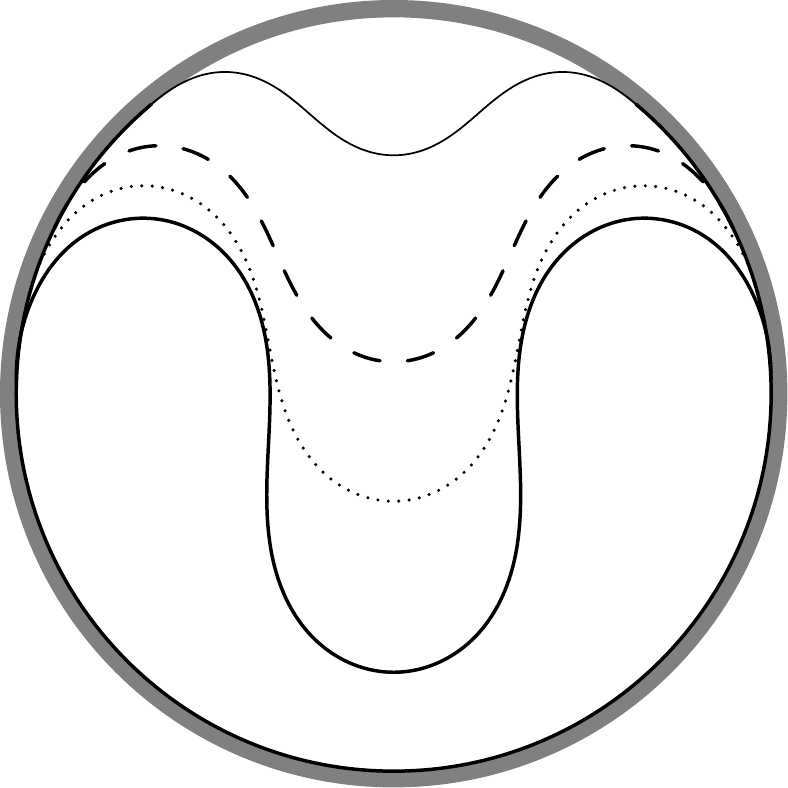} 
\caption{}
\label{fig:shapes_1}
\end{subfigure}
\quad
\begin{subfigure}{0.35\textwidth}
\includegraphics[width=\textwidth]{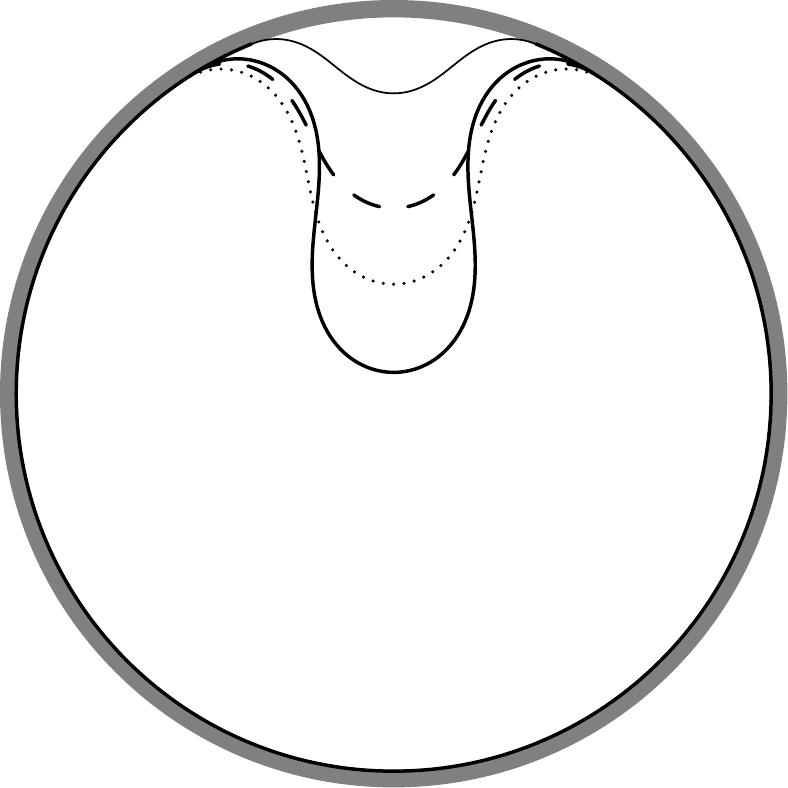}    
\caption{}
\label{fig:shapes_2}
\end{subfigure}
\caption{\small Equilibrium shapes of the elastic strip calculated with (a) zero capillarity ($\rho = 1$) and (b) $\rho=5$, obtained using the  following values for the excess length: $\eps=0.01$ (thin solid lines), $\eps=0.05$ (dashed lines), $\eps=0.1$ (dotted lines) and $\eps=0.2$ (thick solid lines). In absence of elasto-capillarity, Figure (a) shows a decrease of the adherence length for increasing excess-length. By contrast, Figure (b) clearly shows that the adherence length is a non-monotonic function of $\eps$ for moderate capillarity effects ($\rho \gtrsim 1.9$).}
\label{fig:shapes}
\end{figure}

The region of the  $(\eps,\varrho)$-plane in which our solutions are admissible is sketched in Figure \ref{fig:limits}. We gather that, for $\varrho < \varrho^{*}$, with $\varrho^{*}\approx 2.916$, the contact with the wall occurs before the self-contact, and viceversa for $\varrho > \varrho^{*}$. We also find that, while the blister height is an increasing function of the total length, the adherence length may exhibit a non-monotonic behavior. Thus, with reference to  Figure \ref{fig:limits}, $\ell^{(adh)}$ decreases with $\eps$ in the region below the dashed line, while it reverse its behaviour in the region above. Obviously, this change of slope occurs only if the adherence strength is suitably large. This is clearly displayed in Figure \ref{fig:shapes}, where the equilibrium shapes of the elastic strip are plotted for different excess-lengths and for two values of the adherence strength, corresponding to $\rho=1$ (no capillarity) and $\rho=5$.

\subsection{Asymptotic analysis} 
\label{sec:asymptotic}
The main aim of this Section is to provide an approximation to some physically relevant quantities when the length of the strip slightly exceeds that of the confining circumference. To this end, we first derive the approximations of $\bt$ and $\tho$ by performing an asymptotic expansion of Eqs.\eqref{prima} and \eqref{seconda} in the limit $\eps\ll 1$, where $\eps$ is defined as in \eqref{eq:epsilon_rho}. Subsequently, we apply these results to find the approximations of $\ell^{(adh)}$ (complementary to the blister width), $\delta$ and the internal stresses.

\begin{figure}[ht]
\centering
\begin{subfigure}{0.47\textwidth}
\includegraphics[width=\textwidth]{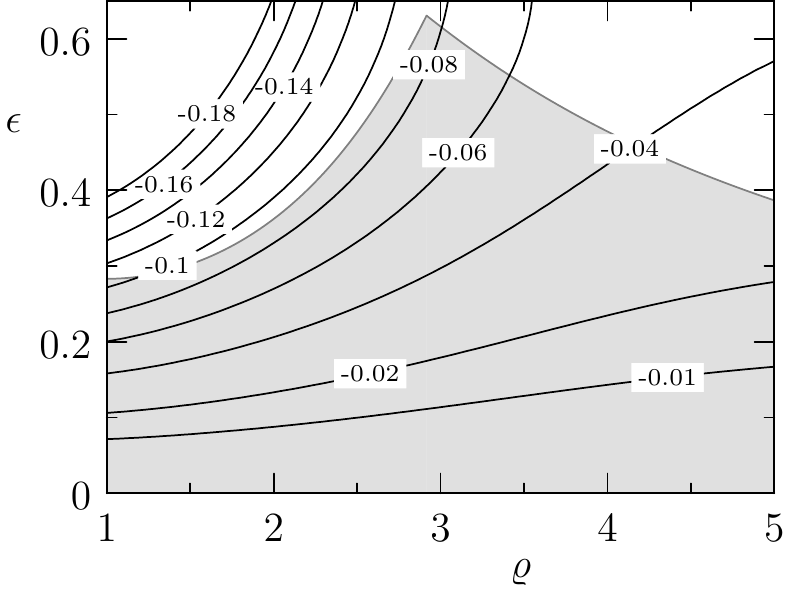} 
\caption{}
\label{fig:relErrLadh}
\end{subfigure}
\quad
\begin{subfigure}{0.47\textwidth}
\includegraphics[width=\textwidth]{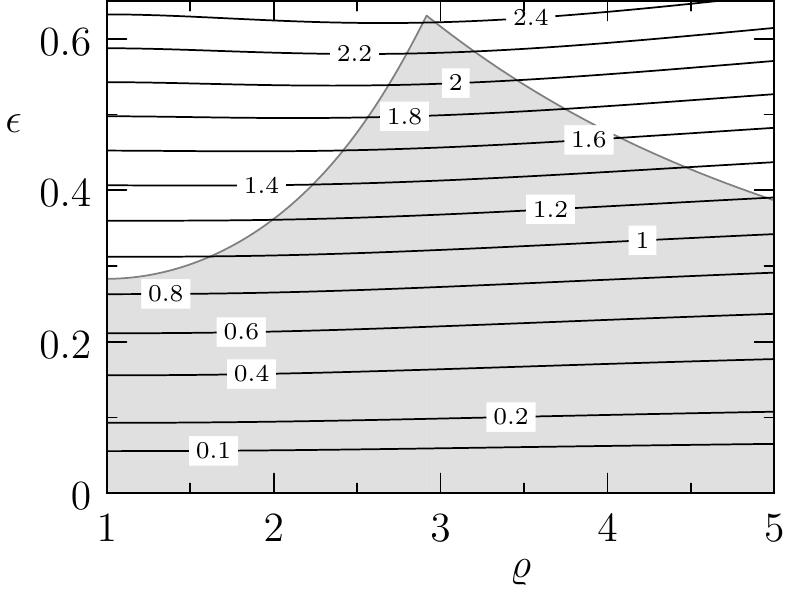}    
\caption{}
\label{fig:relErrDelta}
\end{subfigure}
\caption{{\small Relative-error contour-lines of the two-term approximations to blister height (a) and the adherence length (b). The grey shaded region identifies the limit of validity of our approximation as we neglect the contact with the container at the bottom and the self-contact (see Figure \ref{fig:limits}).}}
\label{fig:relErr}
\end{figure}

When $\eps = 0$, there is only the trivial solution: $\tho=\bt=0$. Since we don't expect any singular behaviour in the solution of the problem at hand, we look for asymptotic expansions where $\tho(\eps) = o(1)$ and $\bt(\eps)=o(1)$, as $\eps\!\downarrow\! 0$. However, we do not make at present any specific assumption on the ratio $v(\eps)=\bt(\eps)/\tho(\eps)$. Next, we substitute the leading approximations for the elliptic integrals, as given in equations \eqref{eq:Fleading}, \eqref{eq:Eleading} and consider only the leading approximation to the following function
\begin{equation}
\sqrt{\frac{\cos\bt-\cos\tho}{1-\cos\tho}} \sim {\sqrt{1-v^2}} \, .
\end{equation}
After a simple manipulation, equations \eqref{prima} and \eqref{seconda}  can then be recast in the following form, 
\begin{subequations}
\begin{align}
\tho \big[ \sqrt{1-v^2} (\pi & - \arcsin v )  + \varrho v \big]  = \pi \eps \varrho \, , \label{eq:eq1_leading} \\
\frac{\tho^3}{12}\big[3 \sqrt{1-v^2} (\pi & - \arcsin v )  + v^3(3-2 \varrho) + v (6\varrho-3) \big] \notag \\
& = -\pi \eps \varrho \left(1 - \frac{\tho^2}{2} + \frac{\tho^4}{24} \right) \label{eq:eq2_leading}
\end{align}
\end{subequations}
which is particularly suited to a dominant balance argument \cite{Bender:1999}. From Eq.\eqref{eq:eq2_leading}, we recognize that the only possible asymptotic balance is \mbox{$\tho \sim a_1 \eps^{1/3}$}, with $a_1$ to be determined. Eq.\eqref{eq:eq1_leading} then implies that the leading order term of $v(\eps)=v_0+o(\eps)$ must satisfy the following equation
\begin{equation}
\varrho v_0 +  \sqrt{1-v_0^2} (\pi - \arcsin v_0 )=0 .
\label{eq:v0}
\end{equation}
These results show how to extend the asymptotic analysis to higher orders. In particular, we know that: $(i)$ the expansion is regular with an asymptotic sequence given by $(\eps^{1/3},\eps^{2/3}, \ldots , \eps^{k/3})$; $(ii)$ $\bt$ and $\tho$ have the same asymptotic behavior, i.e, $v_0=O(1)$. Thus, we simplify the elliptic integrals in Eqs.\eqref{prima} and \eqref{seconda} as described in detail in \ref{app:elliptic} --specifically, we use Eqs.\eqref{eq:F_O4}, \eqref{eq:E_O4}-- and then look for solutions of the following form:
\begin{subequations}
\begin{gather}
\tho  = a_1    \eps^{1/3} + a_2    \eps^{2/3} + a_3    \eps + o(\eps) \, ,\\
\bt  = v_0 \big(a_1    \eps^{1/3} + b_2    \eps^{2/3} + b_3    \eps + o(\eps) \big) \, .
\end{gather}
\label{asy_adhesion}
\end{subequations}
The substitution of these expressions into Eqs.\eqref{prima}, \eqref{seconda} yields the following equations for the coefficients
\begin{subequations}
\begin{align}
a_1    & = \left[\frac{12 \pi \varrho v_0^{-1} }{ 3-3 \varrho +v_0^2 (2 \varrho - 3)}\right]^{1/3} \\
a_2    & = 0 \\
a_3    & = \big[15 \pi \varrho \left(1-v_0^2\right)^2 \left(20 v_0^2 - 1 \right) + 2 \pi  \varrho^2 \left(15-320 v_0^2+497 v_0^4-192 v_0^6\right) \notag\\
& + \pi  \varrho^3 \left(-15+310 v_0^2-384 v_0^4+120 v_0^6\right)\big] f(\varrho)^{-1} \\
b_2    & = 0 \\
b_3    & = \big[15 \pi \varrho \left(1-v_0^2\right)^2 \left(21-2 v_0^2\right) - 2 \pi  \varrho^2 (1-v_0^2) \left(315-225 v_0^2+8 v_0^4\right) 
\notag \\
& +\pi  \varrho^3 \left(315-420 v_0^2+136 v_0^4\right)\big] f(\varrho)^{-1} \, , \\
f(\varrho) & = 40 v_0 (\varrho + v_0^2 - 1) [3-3 \varrho + v_0^2 (2 \varrho - 3)]^2 \, ,
\end{align}
\end{subequations}
where $v_0$ is again given by Eq.\eqref{eq:v0}. Figure \ref{fig:relErr} reports the relative errors of our approximations in the range of interests.

The asymptotic expansion of $\bt$ immediately yields the behaviour of $\ell^{(adh)}$ as a function of $\eps$ (see Eq.\eqref{eq:laderenza}). However, it is slightly more complicated to obtain a good approximation of the blister height, $\delta$. In fact, the Taylor expansion of Eq.\eqref{eq:blister_height}, once $\bt$ and $\tho$ are expressed in terms of $\eps$, only very slowly converge to the numerical solution and thus not provide a good approximation when $\eps$ varies in the range of Figure \ref{fig:relErr}.  However, the two-term approximation is still accurate, within a 10\% of relative error (see Figure \ref{fig:relErrDelta}), when \mbox{$\eps <0.05$}. Despite small, this value of the excess-length already accounts for large deformations suitable to direct measurements, since the blister height is of the order of magnitude of the container radius (see Figure \ref{fig:shapes}). Thus, the two term approximation can be used to compare the theoretical predictions with experiments.

% A possible improvement of the convergence of the series is obtained via the diagonal second-degree Pad\'e approximant \cite{Bender:1999} of Eq.\eqref{eq:blister_height} for $\bt,\tho \ll 1$:
% \begin{align}
% \delta & \sim \frac{6}{\varrho} \left(2 \tho^2 - (2-\varrho) \bt^2 + 2 \tho (\tho^2-\bt^2)^{1/2}\right)^2
% \big[\tho (\tho^2-\bt^2)^{1/2}(24 + 2\tho^2+\bt^2) \notag \\
% & + 24 \tho^2+2 \tho^4 - \bt^2 \left(12+\bt^2\right) (2-\varrho ) \big]^{-1} .
% \label{eq:padedelta}
% \end{align}

We now turn to the discussion of two important limiting cases of adherence: (i) the pure-curvature regime $(\varrho=1)$ and (ii) the elasto-capillarity regime $(\varrho \gg 1)$.  

\subsubsection{Curvature regime ($\varrho =1$)}
\label{sec:subasymptotic_I}
In this regime the only source of adherence is due to the curvature of the confining wall. When $\varrho=1$, equation \eqref{eq:v0} is solved by $v_0=\vc$, with $\vc \approx -0.9761$. The expansion coefficients reduce to
\begin{subequations}
\begin{align}
a_{1}^{_{(1)}} & = -\frac{(12 \pi)^\frac{1}{3}}{\vc}, \label{eq:expa0} \\
a_{3}^{_{(1)}} & = \frac{\pi }{2 \vc}\left(\frac{9}{5} - \frac{1}{4 (\vc)^2} \right), \\
b_{3}^{_{(1)}} & =  \frac{\pi}{4 \vc}\left(-\frac{7}{5} + \frac{9}{2 (\vc)^2} \right) \, .
\end{align}
\end{subequations}
\begin{figure}[ht]
\centerline{\includegraphics[width=0.5\textwidth]{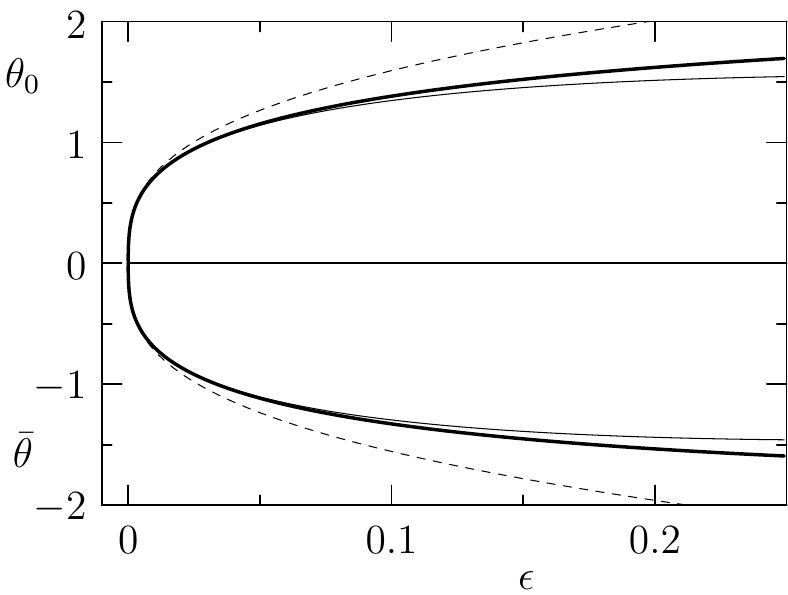}}
\caption{\small Plot of $\bt$ and $\t0$ versus $\eps$ for $\varrho=1$. The thick solid line represents the numerical solution. The one-term approximation (dashed line) agrees with the numerical solution only for very small values of $\eps$. The two-term asymptotic expression (solid thin  line) gives a much better agreement on a wider range of $\eps$.}
\label{fig:theta}
\end{figure}
\begin{figure}[ht]
\centering
\begin{subfigure}{0.47\textwidth}
\includegraphics[width=\textwidth]{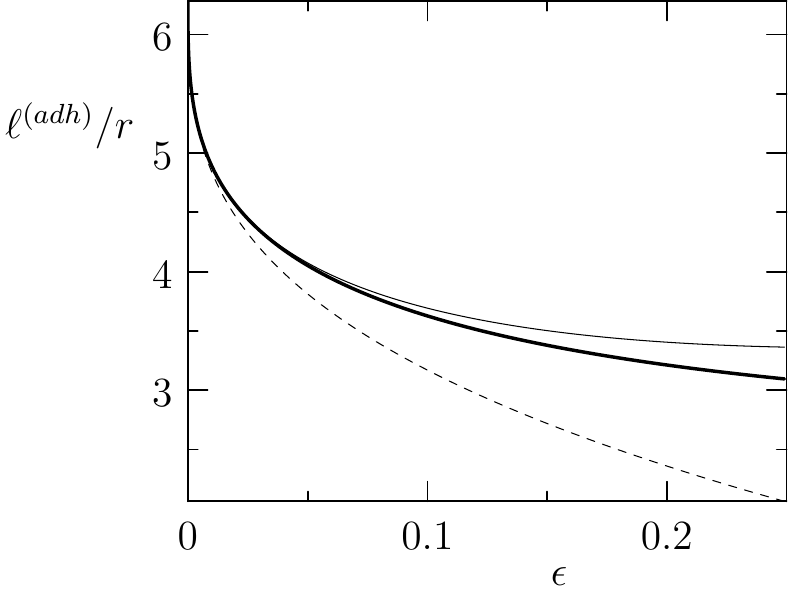} 
\caption{}
\label{fig:NumAsy_Ladh}
\end{subfigure}
\quad
\begin{subfigure}{0.47\textwidth}
\includegraphics[width=\textwidth]{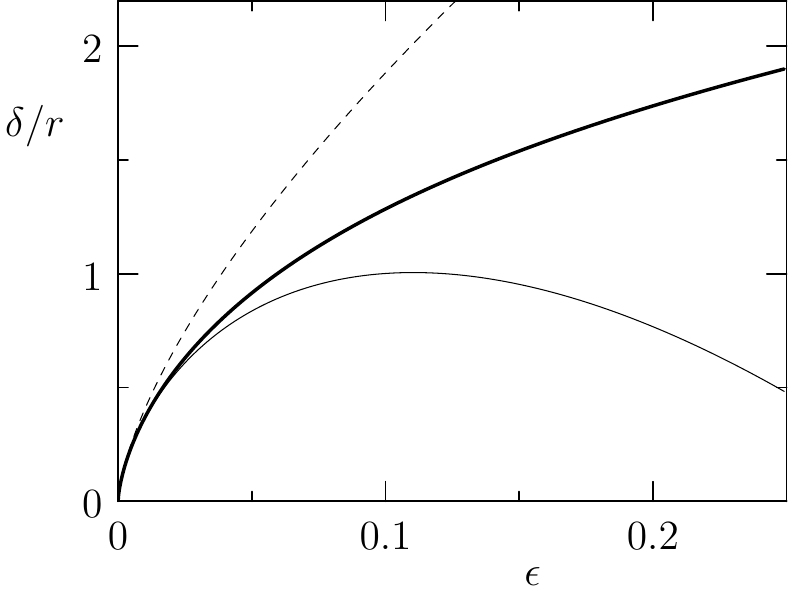}    
\caption{}
\label{fig:NumAsy_delta}
\end{subfigure}
\caption{\small Plot of (a) the adherence length, $\ell^{(adh)}/r$, and (b) the blister height, $\delta/r$, as functions of $\eps$, when $\varrho=1$. The solid thick lines are the numerical solutions, the dashed lines represent the one-term approximation while the solid thin lines show the two-term approximation. This latter approximation seems to describe the adherence length reasonably well, but it does not fully capture the behaviour of the blister height.}
\label{fig:dl}
\end{figure}

Figure \ref{fig:theta} sketches the angles $\bt$ and $\t0$ as functions of $\eps$. The comparison with the numerical approximation clearly shows that the two-term approximation is needed in order to better capture the behaviour of the solution in the whole range of interest. This approximation is also sufficient to describe the adherence length, as shown in Figure \ref{fig:NumAsy_Ladh}. However, as already discussed in \S\ref{sec:asymptotic}, Figure \ref{fig:NumAsy_delta} clearly shows that the blister height is accurately represented by the two-term approximation only in a limited range of $\eps$.
\subsubsection{Elasto-capillarity regime ($\varrho \gg 1$)}
\label{sec:subasymptotic_II}
Whenever the adhesive potential is dominant, we have $\ell_{ec} \ll r$ and, hence, $\varrho \rightarrow \infty$. In this case, equation \eqref{eq:v0} yields
\begin{equation}
\va(\varrho) \approx -\frac{\pi}{\varrho},
\end{equation}
and consequently
\begin{equation}
a_{1}^{_{(\infty)}}(\varrho) \approx 2^\frac{2}{3} \varrho^\frac{1}{3}, 
\qquad a_{3}^{_{(\infty)}}(\varrho) \approx \frac{1}{24} \varrho, 
\qquad b_{3}^{_{(\infty)}}(\varrho) \approx -\frac{7}{8}\varrho.
\end{equation}
It is worth comparing the length of the free part of the strip and the blister height with the analogue quantities in the planar case as given by formulas (9) and (10) of \cite{Wagner:2013}. To this end, we observe that the {\it compression $\Delta l$}  can be express in terms of $\eps$ as
$\Delta l  := 2 \pi r \eps$. By using $\varrho \approx \sqrt 2 r/\ell_{ec}$, the length of the non-adhering portion (measured in unit of elasto-capillarity length) is 
\begin{equation}
-\frac{2 r}{\ell_{ec}} \sin \bt_{\infty}  \approx -2 \frac{r}{\ell_{ec}} \bt_\infty = 2 \pi^{2/3} \left(\frac{\Delta l }{\ell_{ec}}\right)^\frac{1}{3} - \frac{7}{8}\frac{\Delta l }{\ell_{ec}} .
\end{equation}
Similarly, we obtain the expression for the blister height
\begin{equation}
\frac{\delta_\infty}{\ell_{ec}} \approx  2 \sqrt 2 \left(\frac{\Delta l}{ \pi \ell_{ec}}\right)^\frac{2}{3} -  \frac{1}{2\sqrt2}\left(\frac{\Delta l}{ \pi \ell_{ec}}\right)^\frac{4}{3} .
\end{equation}
Not surprisingly, these results coincide with those reported in \cite{Wagner:2013}.
\section{Adherence by curvature with unilateral contact}
Let us now suppose that the container can be modelled as a unilateral and frictionless contact ($w=0$). This means that the wall can exerts only contact forces directed along the inward normal direction. We discuss this problem from the mechanical point of view, within the theory of the Euler-Bernoulli beam. Accordingly, at equilibrium the internal force $\Tv(s)$ and the internal torque $\Mv(s)$ obey the following equilibrium equations
\begin{equation}
\frac{\d \Tv(s)}{\d s} + \fv(s) = {\bf 0}, \qquad \frac{\d \Mv(s)}{\d s} + \tv(s) \times \Tv(s) + \mv(s) =  {\bf 0},
\label{uni}
\end{equation}
where $\fv$ and $\mv$ are the external forces and torques per unit length, respectively. This equations must hold in any section $s\in [s_1,s_2]$ of the curve. Since we assume the effects of gravity to be negligible, the only source of external distributed forces is the contact force exerted by the container, while $\mv={\bf 0}$. In the presence of a concentrated force $\Fv$ and torque $\bGa$ at $s = s_*$ the following local balances hold
\begin{equation}
\lim_{s \to s_*^+} \Tv(s) -  \lim_{s \to s_*^-} \Tv(s) =  \Fv, \qquad \lim_{s \to s_*^+} \Mv(s) -  \lim_{s \to s_*^-} \Mv(s) = \bGa .
\label{salto}
\end{equation} 
This system of equations is completed by the Euler constitutive equation, that, within the hypothesis of plane deformations, states that the internal torque is proportional to the difference between the curvature and the intrinsic curvature $c_0$:
\begin{equation}
\Mv(s) = \kappa[\theta_s(s) - c_0]\ev_z.
\label{eulero}
\end{equation}\\

The free part of the beam is not subject to any external distributed load. Therefore, \eqref{uni}$_1$ shows that $\Tv$ must be constant throughout $s\in[0,\bs)$. The equilibrium equation of the free part is thus provided by the balance of torque \eqref{uni}$_2$ which reads
\begin{equation}
\kappa \theta_{ss} - T_x \sin \theta + T_y \cos \theta = 0,
\label{eq:equilibrium_Euler}
\end{equation}
where $T_x$ and $T_y$ are the Cartesian components of the internal force. We recall that, by our convention, the tangent unit vector to the beam is $\tv(s) = \cos \theta(s) \ev_x + \sin \theta(s)\ev_y$. At first sight, equation \eqref{eq:equilibrium_Euler} differs from the equilibrium equation \eqref{pendolo} as it contains a term in $\cos\theta$. However, we have assumed that $\theta(s)$ is odd (and therefore also $\theta_{ss}$ is odd). It is then easy to show that $T_y$ must vanish and the torque equation reduces to \eqref{pendolo}. As a further consequence of this symmetry, we observe explicitly that from $T_y=0$ it also follows that the constant internal stress, $\Tv$, is purely horizontal: $\Tv=T_x \, \ev_x$.
\begin{figure}[ht]
\centerline{\includegraphics[width=0.6\textwidth]{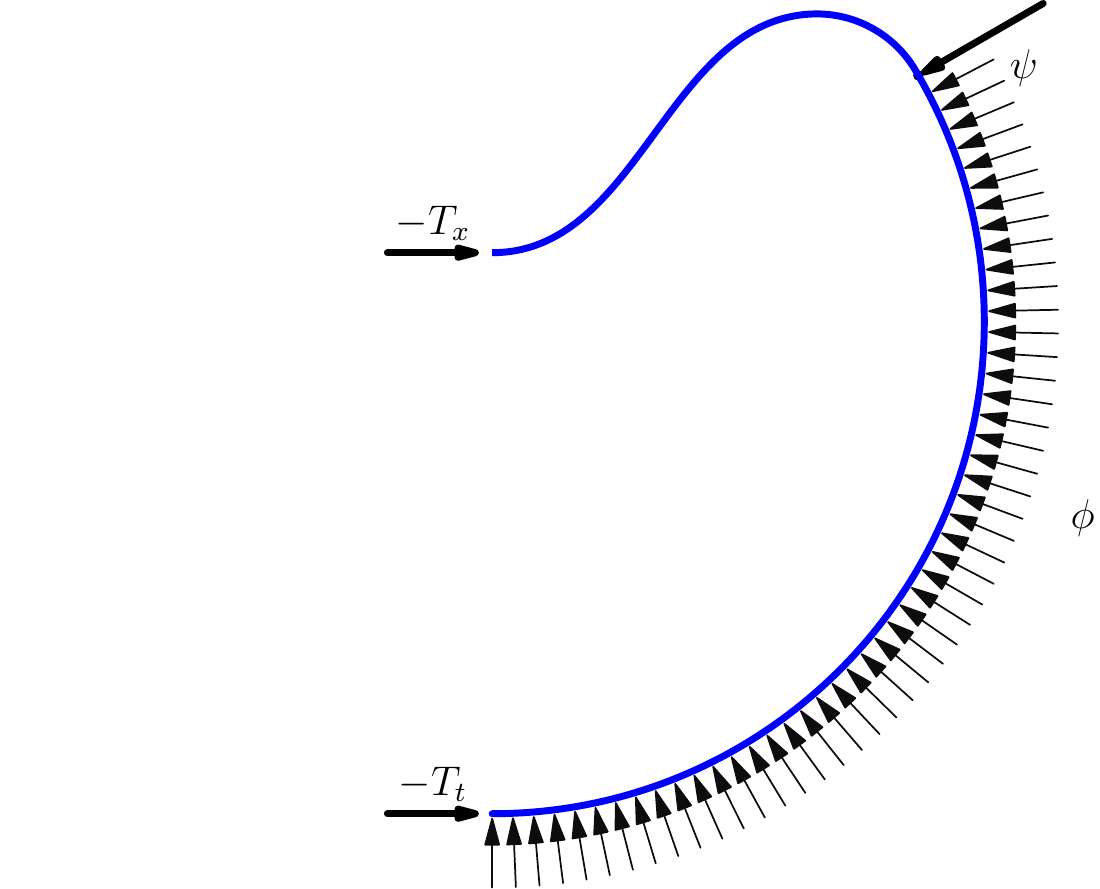}}
\caption{\small Internal and external forces acting on the half-beam in case of unilateral and frictionless contact. The forces $T_x$ and $T_t$, as given in equations \eqref{tt} and \eqref{tx} respectively, are negative, so that the beam is under compression for any admissible configuration. The concentrated force $\bpsi$ is necessary to balance the vertical components of the internal forces at $s=\bs$.}
\end{figure}\\

When the beam is in contact with the external container, the balance of forces requires the introduction of the contact forces, whose density per unit length will be denoted by $\bphi(s)$.  Since we model the container as an ideal unilateral frictionless constraint, $\bphi$ is directed along the inward normal to the surface, hence, we assume $\bphi(s) = -\phi(s) \nv(s)$, with $\phi(s) \ge0$, where $\nv(s) = -\sin \theta(s) \ev_x + \cos \theta(s) \ev_y$. Furthermore, the curvature of the beam in contact with the container is constant and Eq.\eqref{eulero} implies that also the internal moment $\Mv$ is constant.  Thus, from \eqref{uni}$_2$,  we obtain that the normal component of the internal force, $T_n$, must vanish for $s\in(\bs, \ell/2]$. On the other hand, equation \eqref{uni}$_1$, projected along $\nv$ and $\tv$, gives 
\begin{equation}
\phi = -\frac{T_t}{r} , \qquad T_t = \text{constant}
\label{fn}
\end{equation}
where $T_t$ is the axial component  of $\Tv$. This also implies that necessarily $T_t\le0$.\\
 
More subtle is the discussion of the balances at detachment point. To put the problem in the right perspective, we isolate a small portion of beam around the detachment point. Since the internal force in the adherent part posses a non-zero vertical component, while $T_y=0$ in the free part, the balance of forces requires the introduction of a concentrate reactive force $\bpsi = -\psi \nv(\bs)$. This can only be given by the container, and therefore it is necessarily direct as the inward normal $(\psi \ge 0)$. More precisely, the balance in the $y$-direction is 
\begin{equation}
T_t \sin \bt =  \psi \cos \bt. 
\label{tt}
\end{equation}
Thus, $\psi$ is non-negative (and the contact is truly unilateral) only for $\bt \in [-\pi/2,0]$. On the other hand, the continuity of the axial internal force yields $T_t \cos \bt = T_x - \psi \sin \bt $ whence
\begin{equation}
T_t = T_x \cos \bt.
\label{tx}
\end{equation}
The restriction $\bt \in [-\pi/2,0]$ implies that $T_x$ is non-positive, in agreement with the fact that the entire beam is under compression. Finally we remark that (when $w=0$) the transversality condition \eqref{aderenza} implies the continuity of the curvature at $\bs$ and, thus, the continuity of the internal torque.\\

It is now of special interest to study the asymptotic behaviour of the reactive forces $\phi$ and $\psi$. Equations \eqref{fn}, \eqref{tt} and \eqref{tx} yield $\phi = - T_x \cos \bt / r$ and $\psi = T_x \sin \bt$. Recalling that $T_x = -\kappa \tau$, we gather that both $\phi$ and $\psi$ diverge as $\eps$ goes to zero, because $\tau$ diverges. In fact, by using \eqref{eq:tau} together with the asymptotic expansions  \eqref{asy_adhesion} and \eqref{eq:expa0}, to leading order we find 
\begin{equation}
\tau \sim  \frac{1}{r^2 (1-(\vc)^2)\t0^2} \sim \frac{(\vc)^2}{r^2 (1-(\vc)^2)(12 \pi \eps)^{2/3}} .
\end{equation}
Albeit unexpected, this result is in agreement with the experimental results reported in \cite{Boue:2006}, where it is shown that the mean pressure exerted by the strip on the container becomes very large when $\eps$ tends to zero. 

\section{Concluding remarks}
We have studied the morphology of an elastic closed inextensible strip of length $\ell$, confined by a cylinder of radius $r$, where $\ell > 2 \pi r$. The excess length forces the beam to detach from the cylinder, leading to two distinct parts: an adhering  portion and a free part (or   `blister'). These regions are governed by different equations and must agree at the detachment point, whose position is part of the problem. 

Two different mechanisms concur to promote the adhesion. The first is purely geometric and is the curvature of the container. The second has a physical origin and it is given by the elasto-capillarity interaction of the strip with the container. At human length scales the former usually dominates. However, at small scales the elasto-capillarity often plays a significant role in many phenomena \cite{Liu:2012}.

We have presented numerical results for the equilibrium shape when the strip length $\ell$ and the adhesion strength are given, allowing for the possibility of large deformations.  At fixed $\ell$, the solution depends upon a dimensionless parameter $\varrho \in [1,\infty)$ that measures the relevance of the adhesion due to the curvature with respect to that due to the adhesive potential. The geometrical aspects dominate whenever $\varrho$ approaches one. On the contrary, for very large  $\varrho$, while the elasto-capillarity length remains finite, we match the results that apply to the formation of delamination blisters on a rigid flat substrate \cite{Wagner:2013}. 

In addition to the numerical results, we have provided the asymptotic expansions for two quantities related to the blister shape: the length of the adhering region $\ell^{(adh)}$  and the blister height $\delta$. The small parameter used in these expansions is the normalized excess length $\eps := (\ell  - 2 \pi r)/(2\pi r)$.  The two-term approximation is able to capture the behaviour of $\ell^{(adh)}$ up to the points of self-contact or contact of the blister with the delimiting wall. By contrast, the same approximation predicts the blister height accurately only in a smaller range of $\eps$. In any case, the asymptotic analysis yields simple laws that an experimentalist can possibly use to determine some constitutive parameters by inverse analysis.  

Finally, we have considered the case where the delimiting  wall is modelled as an ideal frictionless unilateral contact and hence determined the external actions that the surface exerts on the strip. These consist in a distributed force (per unit length) and, unexpectedly, also in concentrated force acting at the detachment point. The latter makes the derivative of the curvature discontinuous at the detachment point and is also responsible for the discontinuity of the internal shear force. We also find that when the detachment angle, $\bt$, reaches $\pi/2$, the contact force exerted by the container changes sign, thus violating the unilateral constraint. This effect places an upper limit to the value of admissible $\eps$. More precisely, we find that this effect appears for $\eps \approx 0.228$, a value below that attained for the blister contact with the container. Furthermore, our asymptotic analysis has shown that the internal force and, consequently, the external actions diverge when $\eps$ tends to zero.
 This 
agrees with the experimental results reported in \cite{Boue:2006}. In our opinion, the origin of this singularity could be a consequence of the assumed inextensibility. In a more realistic model, one should relax the inextensibility constraint in favor of a penalization energy term related to compression/dilatation. Thus, initially the strip may undergo a slight compression and then form a blister, beyond a compression threshold. 

% As a final comment, we observe that the studied phenomenon is not a bifurcation, in the sense that the configuration, as a function of $\eps$, is not continuous at $\eps = 0$.  For instance, if we compute the curvature at $s=0$, we obtain $\theta_s (0)=-1/r$ when $\eps=0$, while we find
% \begin{equation}
% \theta_s(0) = \frac{\eta}{\sqrt{1 - v^2}} \quad {\rm as} \quad \eps \rightarrow 0^+.
% \end{equation}
% Thus, $|\theta_s(0)|$, which is the maximum value of the curvature, remains finite, while $\eps$ become infinitesimal. \stecomm{non ho capito bene che cosa vuoi dire....} This implies that the total energy is continuous at $\eps=0$. More precisely, a direct calculation shows that the excess of energy $\Delta W$ induced by the excess of length $\eps$ scales as $\eps^{1/3}$. 

\section*{Acknowledgements} The authors would like to thank A. Goriely, L. De Lorenzis and C. Morosi for their fruitful comments and helpful discussions. RDP is grateful to the Engineering and Physical Sciences Research Council for funding this work via grant EP/H050779/1.
GN and SST acknowledge support from the Italian Ministry of University and Research through the Grant No. 200959L72B 004 `Mathematics and Mechanics of Biological Assemblies and Soft Tissues.'

% {\small
% \bibliographystyle{elsarticle-num.bst}
% \bibliography{elastica.bib}}

\begin{thebibliography}{10}
\expandafter\ifx\csname url\endcsname\relax
  \def\url#1{\texttt{#1}}\fi
\expandafter\ifx\csname urlprefix\endcsname\relax\def\urlprefix{URL }\fi
\expandafter\ifx\csname href\endcsname\relax
  \def\href#1#2{#2} \def\path#1{#1}\fi

\bibitem{Williams:1997}
J.~Williams, Energy release rates for the peeling of flexible membranes and the
  analysis of blister tests, International Journal of Fracture 87~(3) (1997)
  265--288.

\bibitem{Wagner:2013}
T.~J.~W. Wagner, D.~Vella, The `sticky elastica': delamination blisters beyond
  small deformations, Soft Matter 9~(4) (2013) 1025--1030.

\bibitem{Rosso:1998}
R.~Rosso, E.~G. Virga, Adhesion by curvature of lipid tubules, Continuum. Mech.
  Thermodyn. 10~(6) (1998) 359--367.

\bibitem{Goriely:2006}
A.~Goriely, S.~Neukirch, Mechanics of climbing and attachment in twining
  plants, Phys. Rev. Lett. 97~(18) (2006) 184302.

\bibitem{Chen:2007}
J.-S. Chen, C.-W. Li, Planar elastica inside a curved tube with clearance,
  International journal of solids and structures 44~(18) (2007) 6173--6186.

\bibitem{Domokos:2003}
G.~Domokos, W.~Fraser, I.~Szeber{\'e}nyi, Symmetry-breaking bifurcations of the
  uplifted elastic strip, Physica D: Nonlinear Phenomena 185~(2) (2003) 67--77.

\bibitem{Boue:2006}
L.~Bou{\'e}, M.~Adda-Bedia, A.~Boudaoud, D.~Cassani, Y.~Couder, A.~Eddi,
  M.~Trejo, Spiral patterns in the packing of flexible structures, Phys. Rev.
  Lett. 97~(16) (2006) 166104.

\bibitem{Liu:2012}
J.-L. Liu, X.-Q. Feng, On elastocapillarity: A review, Acta Mechanica Sinica
  28~(4) (2012) 928--940.

\bibitem{Kahraman:2012}
O.~Kahraman, N.~Stoop, M.~M. M{\"u}ller, Morphogenesis of membrane
  invaginations in spherical confinement, EPL 97~(6) (2012) 68008.

\bibitem{Patricio:1998}
P.~Patricio, M.~Adda-Bedia, M.~Ben~Amar, An elastica problem: instabilities of
  an elastic arch, Physica D: Nonlinear Phenomena 124~(1) (1998) 285--295.

\bibitem{Domokos:1997}
G.~Domokos, P.~Holmes, B.~Royce, Constrained euler buckling, J Nonlinear Sci
  7~(3) (1997) 281--314.

\bibitem{Fomin:1963}
I.~M. Gelfand, S.~V. Fomin, Calculus of variations, Prentice Hall, 1963.

\bibitem{Abramowitz:1970}
M.~Abramowitz, I.~A. Stegun, Handbook of Mathematical Function with Formulas,
  Graphs, and Mathematical Tables, Vol.~55, 1970.

\bibitem{Bender:1999}
C.~M. Bender, S.~A. Orszag, Advanced Mathematical Methods for Scientists and
  Engineers, Springer, 1999.

\end{thebibliography}

\appendix
\section{Asymptotic approximations for elliptic integrals}
\label{app:elliptic}
\newcommand{\uu}{\bt/\tho}
A very simple approximation of $F(x|m)$ for small $x$ can be obtained as follows
\begin{align}
\F(x|m) & = \int_{0}^{x}\frac{dt}{\sqrt{1-m \sin^2 t}} \approx \int_{0}^{x}\frac{dt}{\sqrt{1-m t^2}}
= \frac{1}{\sqrt{m}} \arcsin(\sqrt{m}x) \, , \\
\E(x|m) & = \int_{0}^{x}\sqrt{1-m \sin^2 t} \, dt \approx \frac{1}{2}\Big[x\sqrt{1-m x^2} +\frac{1}{\sqrt{m}}\arcsin(\sqrt{m}x)\Big] \, .
\end{align}
Therefore, the leading approximations of the elliptic integrals contained in Eqs.\eqref{prima}, \eqref{seconda} are found to be
\begin{align}
2\F(\qo) - \F(\qb) & \approx \frac{\tho}{2} \left(\pi - \arcsin(\uu) \right)\, , \label{eq:Fleading} \\
2\E(\qo) - \E(\qb) & \approx \frac{\tho}{4} \Big(\pi - \arcsin(\uu) - \uu \sqrt{1-(\uu)^2}\Big) \, .
\label{eq:Eleading}
\end{align}
Despite quite crude, these approximations are surprisingly good, as some numerical experiments readily show. However, to be on the safe side, we look for more refined approximations. To this end, we adapt the strategy outlined in the electronic supplementary information of Ref.\cite{Wagner:2013}. We make the substitution $u = \sqrt{m} \, \sin t$, $du = \sqrt{m}\, \cos t \, dt$, so that 
\begin{equation*}
dt = \frac{1}{\sqrt{m}} \frac{du}{\sqrt{1-\frac{u^2}{m}}} \, . 
\end{equation*}
We then substitute $m = \csc^2 \tho/2$, and expand for $\tho \ll 1$, 
\begin{equation*}
dt = \sin\frac{\tho}{2} \Big(1+\frac{1}{2}u^2 \sin^2\frac{\tho}{2} + O(\tho^4) \Big) du \, .
\end{equation*}
The incomplete elliptic integrals are then approximated by
\begin{align}
\F(\qb) & \approx \int_{0}^{\frac{\sin \bt/2}{\sin \tho/2}}
  \sin\frac{\tho}{2} \Big(1+\frac{1}{2}u^2 \sin^2\frac{\tho}{2} \Big) \frac{du}{\sqrt{1-u^2}} \, , \label{eq:F_O3} \\
\E(\qb) & \approx \int_{0}^{\frac{\sin \bt/2}{\sin \tho/2}}
  \sin\frac{\tho}{2} \Big(1+\frac{1}{2}u^2 \sin^2\frac{\tho}{2} \Big) \sqrt{1-u^2} \, du \, . \label{eq:E_O3}
\end{align}
These integrals can be computed exactly. However, we are only interested in their approximation for $\tho \ll 1$ and $\bt \ll 1$. After some calculations, which we do not report for brevity, we obtain
\begin{align}
2\F(\qo) - \F(\qb) & \approx \Big(\frac{\tho}{2} + \frac{\tho^3}{96}\Big)\Big(\pi - \arcsin (\uu) \Big) + \frac{\tho^2 \, \bt}{96} \sqrt{1-(\uu)^2} \, , 
\label{eq:F_O4}\\
2\E(\qo) - \E(\qb) & \approx \Big(\frac{\tho}{4} - \frac{\tho^3}{384}\Big)\Big(\pi - \arcsin (\uu) - \uu \sqrt{1-(\uu)^2}\Big) \notag \\
& - \frac{\tho^2 \, \bt}{192} \big(1-(\uu)^2 \big)^{3/2}\, .
\label{eq:E_O4}
\end{align}

\end{document}